\documentclass[10pt,conference]{IEEEtran}
\IEEEoverridecommandlockouts

\usepackage{graphicx}
\usepackage{amsmath}   % required for align environment
\usepackage{mathtools}  % imports amsmath and then some
\usepackage{amsfonts}  % required for mathbb font
\usepackage{url}     % used for typesetting urls, incl. in bib
\usepackage{color}     % good for putting notes in red and blue
\usepackage{hyperref} % makes all citations and references clickable, brining you to the referenced part of the paper.  Also allows html and mailto buttons. 

\newtheorem{psuedo-code}[subsection]{}

\title{PACE: Pattern Accurate Computationally Efficient Bootstrapping for Timely Discovery of Cyber-Security Concepts \thanks{The Department of Homeland Security sponsored the production of this material under DOE Contract Number DE-AC05-00OR22725 for the management and operation of Oak Ridge National Laboratory.} \thanks{This manuscript has been authored by UT-Battelle, LLC, under Contract No. DE-AC05-00OR22725 with the U.S. Department of Energy. The United States Government retains and the publisher, by accepting the article for publication, acknowledges that the United States Government retains a non-exclusive, paid-up, irrevocable, world-wide license to publish or reproduce the published form of this manuscript, or allow others to do so, for United States Government purposes.}}

\author{
\IEEEauthorblockN{Nikki McNeil}
\IEEEauthorblockA{Department of Mathematics\\
	University of Maryland, Baltimore County\\
	Baltimore, MD\\
	\href{mailto:nmcneil1@umbc.edu}{ncmcneiL1@umbc.edu}}\\[.5cm]   %<------ Line breaks in the current column
\IEEEauthorblockN{Bogdan Czejdo}
\IEEEauthorblockA{Department of Computer Science\\
	Fayetteville State University\\
	Fayetteville, NC\\
	\href{mailto:bcezjdo@uncfsu.edu}{bcezjdo@uncfsu.edu}}
\and
\IEEEauthorblockN{Robert A. Bridges}
\IEEEauthorblockA{Computational Sciences\\
 and Engineering Division\\
	Oak Ridge National Laboratory\\
	Oak Ridge, TN\\
	\href{mailto: bridgesra@ornl.gov}{bridgesra@ornl.gov}}\\[.1cm]
\IEEEauthorblockN{Nicolas Perez}
\IEEEauthorblockA{Department of Computer Science\\
	North Carolina State University \\
	Raleigh, NC\\
	\href{mailto: neperez@ncsu.edu}{neperez@ncsu.edu}}
\and
\IEEEauthorblockN{Michael D. Iannacone}
\IEEEauthorblockA{Computational Sciences\\
and Engineering Division\\
	Oak Ridge National Laboratory\\
	Oak Ridge, TN\\
	\href{mailto:iannaconemd@ornl.gov}{iannaconemd@ornl.gov}}\\               
\IEEEauthorblockN{John R. Goodall}
\IEEEauthorblockA{Computational Sciences\\
and Engineering Division\\
Oak Ridge National Laboratory\\
	Oak Ridge, TN\\
	\href{mailto:jgoodall@ornl.gov}{jgoodall@ornl.gov}}
}

\begin{document}
\maketitle
\begin{abstract}  
 Public disclosure of important security information, such as knowledge of vulnerabilities or exploits, often occurs in blogs, tweets, mailing lists, and other online sources significantly before proper classification into structured databases.  
 In order to facilitate timely discovery of such knowledge, we propose a novel semi-supervised learning algorithm, PACE, for identifying and classifying relevant entities in text sources. 
 The main contribution of this paper is an enhancement of the traditional bootstrapping method for entity extraction by employing a time-memory trade-off that simultaneously circumvents a costly corpus search while strengthening  pattern nomination, which should increase accuracy.  
 An implementation in the cyber-security domain is discussed as well as challenges to Natural Language Processing imposed by the security domain.  
\end{abstract}

\section{Introduction}  
\label{intro}
This paper introduces PACE, a novel bootstrapping algorithm for entity extraction, and an application to  cyber-security where domain concepts involving vulnerabilities and exploits are learned from public text sources. 
Often vulnerabilities and exploits are discussed in a variety of obscure yet publicly accessible websites such as mailing lists, blogs, and twitter feeds, long before proper classification into well-known, commonly referenced databases such as the National Vulnerability Database (NVD), Common Vulnerability Enumeration (CVE), Open Source Vulnerability Database (OSVBD), Exploit-DB, and also before vendor patches or mitigations are released\footnote{\url{http://nvd.nist.gov/}, \url{http://cve.mitre.org/}, \url{http://www.osvdb.org/}, \url{http://www.exploit-db.com/}}.  
As this valuable information is often buried in the world-wide web, our overall goal is to automatically obtain this knowledge by extracting  entities from appropriate text sources, with a target audience of security analysts.  
   
While supervised methods for identifying and classifying entities have experienced very accurate results,  this paper explores a semi-supervised technique, as no  labeled training data in the cyber-security domain is available.     
In order to ensure the appropriate concepts are learned, semi-supervised entity extraction, which almost exclusively is some form of bootstrapping \cite{nadeau2007survey} is used with a small hand-labeled training set.  
In particular, our algorithm, PACE, modifies the traditional bootstrapping approach by storing contextual information with known entity names.  

The benefits of this new technique are multi-fold.  
Specifically, as patterns are only learned from the contextual instances observed with known entities, PACE allows more accurate pattern nomination than previous bootstrapping methods.   
Secondly, it obviates the need for extremely large corpora, allowing PACE to be deployed in an operational setting where documents are streamed into the corpus under analysis and are discarded from the corpus after a fixed time.  
Lastly, PACE uses a time-memory trade-off to circumvent a traversal of the corpus previously necessary for pattern nomination.  

% DO I NEED TO SAY EXPLICITLY THAT WE WILL PROCESS DOCUMENTS DAILY AND PROCESS THEM FOR APPROPRIATE ENTITIES - PROBABLY SOMEWHERE!
% probably ok to just imply it will be done often, which the above hopefully does.
%
%\subsection{Outline}
%Section \ref{example} discusses disclosure of vulnerability and exploit information, in particular, their manifestations into online sources.  A brief summary of previous bootstrapping occurs in Section \ref{background}  and is complimented by an introduction to PACE bootstrapping in Section \ref{approach}  along with the implementation details of a prototypical run.  Obstacles to entity extraction which are not stereotypical, but rather  encountered when working in the cyber-security domain are itemized in Section \ref{challenges}, and some directions of future work is the final Section, \ref{future}.

\section{Textual Evidence of Vulnerabilities}
\label{example}

\begin{figure*}
\centering
\setlength\fboxsep{.25pt}
\setlength\fboxrule{0.25pt}
\includegraphics[width=5.7in]{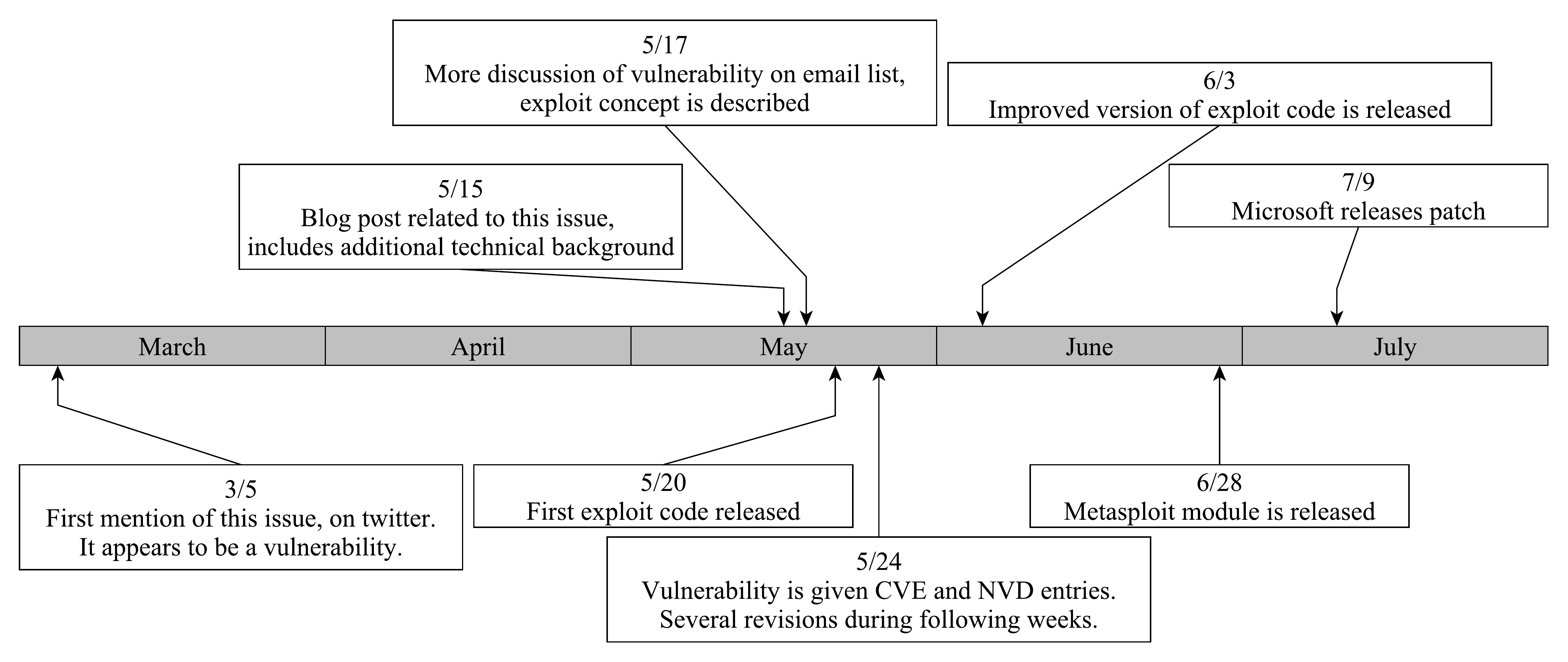}
\caption{Timeline of publicly available information on vulnerability CVE-2013-3660 and related events.}
\label{timeline}
\end{figure*}

As a driving example, we consider the following observed sequence of events where a vulnerability and proof of concept exploit is discovered and discussed publicly online months before proper classification into the aforementioned databases (see Figure \ref{timeline}). 
In this case, a potential vulnerability in Microsoft Windows is discovered and publicly disclosed on Twitter and on a public email list of vulnerability researchers.  
This vulnerability is confirmed by others, a proof of concept exploit is released, and improved versions of this exploit code follow.  
The vulnerability is added to all public vulnerability databases after some delay; these entries are revised many times as the situation develops and new information becomes available.  
Similarly, the exploit code is added to Exploit DB and is incorporated into the Metasploit framework.  
The vendor later releases a patch and related security bulletin as part of their normal patching process.

More generally, disclosure of software vulnerabilities falls into three categories: public ``full disclosure,'' private ``coordinated disclosure,'' or non-disclosure.  
Full disclosure involves the researcher making information public as it becomes available, like the case above; coordinated disclosure involves informing the vendor (e.g. Microsoft) who often will disclose the vulnerability only after a patch is available.
In the remaining cases, the knowledge of the vulnerability can be kept private or sold to a third party.  
The public and the vendor will not learn of it until it is either used in a (zero-day) attack or it is discovered independently by another researcher.  
There is significant debate about the merits and ethics of these approaches, and many intermediate approaches occur in practice.  
Any of these methods may or may not include related exploit code, generally as a proof of concept.

In any of these cases, automatically identifying the relevant entities in online sources will provide more timely information to security analysts, although the specifics of what, when, and where the information is learned will of course vary.  Also, in each case, the same observations generally apply:
\begin{itemize}
  \item Information usually appears in unstructured sources earlier than in structured sources.
  \item There is often a discussion related to each event, and it is often dispersed across several locations. 
(eg. a group of related emails, tweets, Reddit comments, and blog posts.)  This discussion often adds important details, such as the vulnerability's expected impact in practice.
  \item Typically there is no single source for all of the relevant information.
\end{itemize}
As other such examples have been observed, our hypothesis is that much useful information may be discovered in a more timely fashion with the development of natural language processing tailored to the security domain.

\section{Background}
\label{background}
Previous work in the intersection of Natural Language Processing for understanding cyber-security concepts has been undertaken.  In \cite{finin2011wikitology} a combination of databases, Wikipedia, and ``off-the-shelf'' tools are  used to identify and classify vulnerability entities.  
Very recent work of \cite{joshiextracting} address supervised learning for entity extraction in cyber-security, by hand-labeling a small corpus of training data and using an ``off-the-shelf" entity recognizer.  Our efforts also include a supervised approach, but we focus only on bootstrapping here.  
 
\subsubsection{Bootstrapping Techniques for Entity Extraction}
\begin{figure}
\includegraphics[width=3.4in]{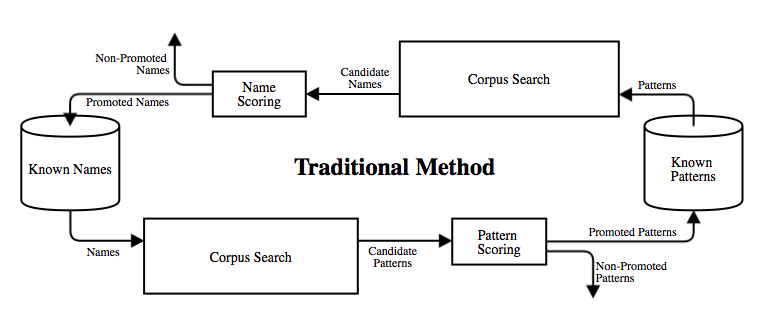}
\caption{A cycle in traditional bootstrapping involves 2 traversals through the corpus, one to nominate new patterns, one to nominate new entity names.}
\label{traditional}
\end{figure}

Almost all semi-supervised techniques for entity extraction use a bootstrapping technique \cite{nadeau2007survey}, and follow a similar overall cyclic structure.  Given an entity type (such as ``president'' or ``vulnerability'' ) a bootstrapping algorithm requires a set of known entity names, a set of known patterns (this is the usually small training set referred to as ``seeds''), and a text corpus (usually large).
A \textit{pattern} is contextual information which gives evidence for identifying a segment of text as an instance of an entity.  
For example, a pattern for identifying presidential names could be a proper noun directly followed by the words ``was inaugurated''.  
Traditionally, bootstrapping searches the corpus for known patterns to produce candidate entity names, which are then scored so only the most trusted names are promoted to join the set of known entity names.  
The corpus is also searched for instances of known entity names and candidate patterns are nominated from the observed context.  
Candidate patterns are then scored to determine promotion.  
This cycle may continue many times for a given corpus, as new patterns and entity names may be learned on each cycle \cite{carlson2009nell, brin1999extracting, carlson2010coupled, riloff2012event, jones2005learning}.  
A diagram of the process is provided in Figure \ref{traditional} .  
Although outside the scope of the current paper, it is commonplace for this algorithmic setup to be implemented for relation extraction, often simultaneously with entity extraction \cite{agichtein2001snowball, carlson2009nell, brin1999extracting, carlson2010coupled}.

\begin{figure}
\includegraphics[width=3.4in]{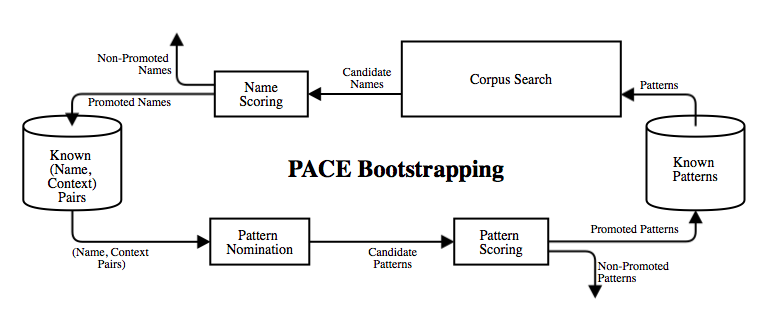}
\caption{A cycle in the new algorithm involves one traversal through the corpus for nominating (entity name, context) pairs.  Storing entity names with their observed context facilitates more robust pattern selection without a second corpus search.}
\label{our_chart}
\end{figure}

While all previous bootstrapping algorithms for entity extraction follow the general workflow discussed above, variations in implementation details have yielded worthwhile results. 
In \cite{carlson2010coupled}, a predetermined ontology of entity and relation types is used to impose constraints on the learned instances that results in greater accuracy.  Active learning can be incorporated by periodically requesting human feedback  in order to omit spuriously learned patterns and entities, as such drifting is a common problem for bootstrapping techniques~\cite{carlson2010active}. 

The advantage of bootstrapping is the minimal required labeled data, which facilitates its use in almost any domain.  
On the other hand, the perennial Achilles' heel of bootstrapping, and more generally of any machine learning with minimal training data, is the acquisition of spurious results, causing extracted terms to drift from the desired entity type.  
Traction is gained in the details of the scoring algorithms and pattern selection.   
We note that the overall goal of many previous implementations of bootstrapping is to create a comprehensive list of names for a given type; for example, see \cite{carlson2009nell, brin1999extracting, carlson2010coupled, carlson2010website,   riloff2012event}.
  Consequently, if a known entity name is overlooked in one document but found in another, the desired outcome is still accomplished.  On the other hand, if concepts to be learned are dissimilar (e.g. including both ``president'' and ``athletic team'' as entity types), disparate text sources are necessary in the corpus (e.g. both sports news and historical documents), leading to semantic drift. 

A large corpus is generally assumed, sometimes on the order of tens of millions of documents, and, in fact, relied upon.  For example, in \cite{brin1999extracting} extremely stringent rules are imposed when nominating patterns so that only precise patterns are learned.  
While this may reduce drift, recall suffers and the massive size of the corpus is needed for the system to learn anything.  An additional limitation of such an approach is the computational and temporal cost incurred.  
Brin's \cite{brin1999extracting} corpus included more than 24 million documents, on which no implementation of the algorithm was reported completed, and a smaller corpus of approximately 5 million documents took a few days to complete a cycle.  

Considering the motivating example discussed in Section~\ref{example}, the current goal is ideally to identify each occurrence of a security entity in a document for timely discovery of new vulnerabilities and exploits and  to store this information in a database;  
hence, this problem is more of a labeling task instead of creating extensive lists of known entities.
 This difference requires greater recall for each document.  
While Section~\ref{challenges} discusses the many challenges imposed by the complexity of the entities in the security domain, working in exclusively one domain is an advantage;  namely, considering only relevant documents will inhibit drift, which may be accomplished by using a decision classifier to discard irrelevant documents  when populating the corpus, as in \cite{finin2011wikitology}.  
%PACE provides a new algorithm which should increase pattern accuracy, thereby retaining recall in the midst of a small corpus, and is expected to lessen computational expense.

\section{The PACE Bootstrapping Algorithm}
\label{approach}
As cited above, semi-supervised techniques for entity extraction have almost exclusively involved the bootstrapping work flow described in Section \ref{background}.  The main contribution of this paper is  a novel algorithm which enhances the pattern nomination process.  Specifically, by storing known entities along with their respective context, candidate  patterns are generated only from the context surrounding known (i.e. seeded or learned) entities.   Perhaps more importantly, we expect PACE to lessen computational cost, as no corpus search is necessary for nominating patterns.  The algorithm for a fixed entity type is outlined below, followed by a description and is depicted in Figure \ref{our_chart}.\\
\indent{\bf PACE Algorithm}
\begin{enumerate}
%\label{psuedo-code}
\item {\bf Initialize:} Initiate a ``Known Patterns'' set and a ``Known [Entity, Context]-Pairs'' set and populate both with seeds. Set a number of iterations, $n$.
\item {\bf Learn Patterns from Known [Entity, Context]-Pairs:} 
	\begin{enumerate}
	\item {\bf Pattern Nomination:} Compare known [entity, context]-pairs to nominate new candidate patterns.  
	\item {\bf Pattern Scoring and Promotion:} Score candidate patterns and promote highest ranking to join the ``Known Pattern'' set.
	\end{enumerate}
\item {\bf Learn New [Entity, Context]-Pairs:}
   \begin{enumerate} 
	\item {\bf [Entity, Context]-Pair Nomination:} Search the corpus for known patterns to nominate new candidate [entity, context]-pairs.
	\item {\bf [Entity, Context]-Pair Scoring and Promotion:} Score candidate [entity, context]-pairs and promote highest ranking to join the ``Known [Entity, Context]-Pairs'' set. 
	\end{enumerate}
\item {\bf Iterate:} Repeat 2 \& 3 $n$ times. 
\end{enumerate}
%\end{psuedo-code}

%\subsubsection{Advantages of PACE}
%\label{advantages}
Similar to traditional bootstrapping, PACE takes seed patterns, seed entity names, and a text corpus as input, but the seed entity names must come paired with the context in which they were observed, similar to the recently released Google Relation Extraction Corpus\footnote{\url{http://googleresearch.blogspot.com/2013/04/50000-lessons-on-how-to-read-relation.html}}. %\cite{google_relation}.  
Operationally, two sets, one of patterns and one of [entity, context]-pairs are initialized (seeded) and expanded similar to usual bootstrapping.  
Hence, the same name may appear multiple times, each with different context in the seed [entity,context]-pair set.  For preliminary testing, context is defined as a radius of five words; that is, a known entity name is stored with a five token prefix and five token suffix, e.g. Table \ref{table_examples}.  While it may be argued that PACE necessitates an increase in training data, seed patterns are generally manually chosen from a familiar contextual observation of an entity name; therefore, this requirement amounts to storing a handful of ten token context strings, a trivial increase in storage, as seed sets are typically small.  

The immediate boon of adding context to entity names is a strengthening of the  pattern nomination process, as only context of known (i.e. seeded or promoted) entity names is used in learning patterns.  
Because we are limiting the field from which patterns candidates are learned to only relevant contextual examples, less stringent rules for pattern nomination (than previously used) yields worthwhile patterns and increases recall.  
Moreover, this uniquely positions PACE to work effectively on small corpora.    

By contrast, traditional bootstrapping learns pattern candidates by first identifying {\it any} instance of a known entity name in the corpus, and it nominates a pattern from the context regardless of whether the matching text string is actually in reference to the desired entity type or not.  
While this may be plausible (and useful, e.g. see {\it name}-patterns in \ref{patterns}) for entity names such as the president ``Abraham Lincoln'' or the vulnerability effect ``remote code execution,'' which are specific enough to almost always appear in reference to the same entity, when considering more ambiguous names such as ``applications'' (a common reference to software among other uses), relying on the premise that every matching string is indeed an allusion to a specific entity is na\"{i}ve and can quickly lead to learning spurious results.  
Previous attempts to combat this (e.g. see Brin  \cite{brin1999extracting}) is to nominate only extremely specific patterns, which can severely limit recall and then compensate by using an enormous corpus.  
Since their goal is to create a list of entity names, so long as each name is identified once in the corpus, their results do not suffer.  
The reliability on both specific patterns and a large corpus is emphasized in Brin \cite{brin1999extracting}, as is the computational cost where only a fifth of the corpus is traversed in a number of days.   
As discussed in the Results Subsubsection \ref{results}, we tested the necessity of a large corpus with very strict pattern nomination rules and employing one of Brin's  pattern nomination limitations with a very small corpus yielded the PACE algorithm to learn no new patterns.  
In short, as patterns are learned from a trusted set of contexts, less stringent rules on pattern formation are necessary, which produces greater recall and facilitate using a smaller corpus than previously possible. 

While PACE is well-suited to handle small corpora, we expect it's performance to be superior on any sized corpus.  
Since pattern candidates are nominated from the known [entity, context]-pairs,  the need to search the corpus is circumvented during pattern nomination.  
As only one traversal of the corpus (for finding new [entity, context]-pairs) is needed in PACE's cycle,  the computational time should be approximately halved; hence, many of the limitations of previous methods are overcome by PACE.  

\subsection{Implementation Details}
Here we describe the algorithmic details used in a prototypical implementation of PACE. 
We expect the technicalities to be tuned to specific applications to increase performance and we provide the details for completeness and as a starting point.  
To obtain seeds, we hand annotated ten documents from online sources\footnote{\url{http://www.computerworld.com/s/topic/85/Malware+and+Vulnerabilities}, \url{https://community.rapid7.com/community/infosec/blog/2013/06/07/keyboy-targeted-attacks-against-vietnam-and-india}, \url{https://groups.google.com/forum/\#!forum/rubyonrails-security}, \url{http://seclists.org/fulldisclosure}}. 
%\cite{computerworld,rapid7keyboy, RoRsecurity, seclists})  discussing vulnerabilities and exploits.
Each instance of the following four entity types: {\it Exploit Effect, Software Name, Vulnerability Potential Effects, Vulnerability Category}, are labeled and manually extracted along with their five-word prefix and five-word suffix as seed [entity,context]-pairs. 
Twenty additional entity types were labeled in the seeding process but occurred too rarely to produce new instances in practice; hence, a larger seed set will be necessary for these types.  
%A small corpus of ten (different) documents are chosen from \cite{computerworld}.    
\subsubsection{Entities \& Patterns}
\label{patterns}
Examples of [entity, context]-pairs are given in Table \ref{table_examples}.            
Unlike entity types used in previous bootstrapping, many of the names required by this domain are short phrases, and are discussed more in Section \ref{challenges}.  
   
\begin{table*}
    \begin{center}
      \caption{Entity, Context Pair Examples}
		\label{table_examples}        
        \begin{tabular*}{.975\textwidth}{|c|c|c|c|}
        \hline
           	Type	 & Name & Prefix & Suffix\\ \hline
			Vulnerability      & ``bug''  &  ``the malware also abuses a''  & ``in the way Android processes''   \\ \hline
			Vulnerability Potential Effects & ``inject arbitrary PHP code'' & ``that could allow attackers to'' &``and execute rogue commands on'' \\ \hline
			Software Name & ``Android'' & ``exploits previously unknown flaws in'' & ``and borrows techniques from Windows'' \\ \hline
			Vulnerability Category & ``exploitable for remote  & ``unlikely that this vulnerability is'' & ``due to technical constraints'' \\
								&   code execution''& &\\ \hline
        \end{tabular*}		
     	\end{center}
\end{table*}
Patterns can be any combination of prefix, name, suffix strings to be matched in the corpus.  
More specifically, a pattern will be denoted {\it [prefix, name, suffix]} where the {\it prefix (suffix)} is up to five tokens immediately preceding (following)  an entity instance, and the {\it name} is up to ten tokens matching an entity instance.   
In practice, words are stemmed using Python's Natural Language Toolkit \footnote{\url{http://nltk.org/}} \cite{bird2009natural}, and select stopwords including punctuation tokens are removed before the strings are matched.  
The trivial pattern, with all three components ({\it prefix, name, suffix}) empty, is discarded (as it would match anything), but, in general, any other combination is allowed. For example, {\it ``allow attacker to \underline{\phantom{blah bh}}'} is a seed pattern for identifying the Vulnerability Potential Effects entity type by matching a three token prefix.
To our knowledge, the use of {\it name}-only patterns, in which the {\it prefix} and {\it suffix} are empty, is novel to bootstrapping and allows quick identification of those entity names which are very specific (such as the Vulnerability Category name {\it ``code injection''} which was also a seeded pattern).  Through both scoring and nominating patterns from only the known entity context, PACE seeks to incorporate name matching in specific cases and still preserve pattern specificity. 

\subsubsection{Nomination of Entities \& Patterns}
Because entity names are often a sequence of words, before a potential entity name can be nominated, chunking (identifying short noun phrases called ``chunks'') is used to identify appropriate phrases for names within the corpus.    After an appropriate name chunk is identified in the corpus, 
%(ending in a noun or of length ten tokens), %% not sure we need this- does it add anything to the paper? I think it will confuse the audience and exposes some not very sophistocated aspects we used 
given a pattern {\it [prefix, name, suffix]}, we require the {\it prefix} and {\it suffix} to exactly match the corresponding tokens preceding and following the name chunk.  Additionally, the name chunk itself must contain the {\it name}.  If these conditions are satisfied, the name chunk, along with its context (five token prefix and suffix), is extracted as a potential entity.
Extracting chunks as entity names is designed to err in favor of longer phrases being extracted for two reasons.  First, these phrases provide greater descriptive accuracy which benefits a security analyst in cases where single words are ambiguous.  Second, the more specific a known entity name is, the less drift the bootstrapping algorithm will encounter.

Pattern nomination is accomplished by comparing known [entity, context]-pairs within an entity type.  
Specifically, if {\it prefix$_1$, name$_1$, suffix$_1$,} 
and {\it prefix$_2$, name$_2$, suffix$_2$}
are two known [entity, context]-pair instances (five token prefix, at most ten token name, and five token suffix) of the same type, we construct a new pattern {\it[prefix, name, suffix]} as follows:  
To define {\it prefix (suffix)}, take the longest matching string between {\it prefix$_1$} and {\it prefix$_2$} ({\it suffix$_1$} and {\it suffix$_2$}) beginning from the right (left); i.e., the string matching begins near the entity name and works outward.  Similarly, {\it name} is defined by matching strings from right to left as generally the end of a name phrase is most informative. Most name phrases end in nouns, which are the most important word in understanding the phrase (similar to a headword).   If at least one of {\it prefix, name,} or {\it suffix} is non-empty, the resulting pattern is nominated.  For large corpora and/or seed sets,  comparison of more than two [entity, context]-pairs is possible.

\subsubsection{Scoring}
Nominated candidates are collected, scored, and ranked, so that the top 50\% of entities (with their respective contexts) and 25\% of patterns are promoted within each type.  Ideally, future testing will yield the promotion percentile to be chosen as a function of the inputs. 
In \cite{carlson2010active}, a study of many scoring techniques is discussed, and the Basilisk method outperforms the others, especially when human intervention is included; consequently,  we employed the Basilisk scoring method for both candidates entities and patterns. 

In order to score a given a candidate entity $e$, let $p_1, ... , p_n$ be the different patterns that matched and nominated $e$ during the corpus search, and let $f_j$ denote the number of known entities pattern $p_j$ has previously matched. 
Then the Basilisk enity score $=\sum_{j=1}^n \frac{\log(1+f_j)}{n}.$  
In traditional bootstrapping, Basilisk scoring provides a separate function for pattern scoring that relies on the ratio of the number of times the pattern is observed with an (assumed) entity name to the total number of times the pattern occurs in the corpus.  
Specifically, given pattern $p$, if $n$ is the number of times the pattern is observed with a string matching the entity name in the corpus, and $N$ is the number of times the pattern is occurs in the corpus, the 
Basilisk pattern score $=\frac{n}{N}\log(n).$  
 Since PACE circumvents the costly corpus search by nominating patterns only from those context instances observed with known entities, we calculate $n$ and $N$ as counts relative to the set of [entity, context]-pairs. 
 The ratio $n/N$ is commonly equal to one in our prototypical experiments.   

\subsubsection{Results}
\label{results}
Initial tests of a PACE bootstrapping prototype have been performed on a small corpus of seven cyber-security news articles\footnote{\url{http://www.computerworld.com/s/topic/85/Malware+and+Vulnerabilities}}, 
%\url{http://seclists.org/fulldisclosure/} 
%from \cite{computerworld}
and corpus documents were disjoint from the documents used to produce seeds.
 Bootstrapping was iterated until a  cycle extracted no new entity names (six iterations). 
 Twenty-three patterns were learned and promoted, of which three extracted inaccurate entities due to their generality, fourteen were accurate but too specific to extract many entities, and six were both accurate and useful in extracting many entities. 
 Twenty-one entity name phrases were extracted from the articles and promoted, of which nineteen were accurate, yielding a precision of 90\%, but a recall of only 12\%.  Omitting entity types for which only a few seeds existed provided a recall as high as 38\%. Clearly a larger seed set will improve results, and more testing is needed to see PACE's potential.  

% Calculating recall varied greatly as discrepancies among scorers, for example, on how long a name phrase should be, and we do not value recall as an accurate metric.
 % captures, and because of decisions about whether every mention of a certain entity name should be labeled within a document if it is not being used in a way to describe its entity type; therefore, we do not value our calculation of recall as an accurate metric here.

Part of Brin's \cite{brin1999extracting} scoring method is to disregard any patterns occurring in different entity names belonging to the same document.  As discussed above, this inhibits drift but also recall.  Testing this rule on the corpus of seven documents yielded zero learned patterns. Once again, Brin's bootstrapping relies on an extremely large corpus, whereas our goal is to be able to extract as much information as possible from even a single document.

\section{Challenges Imposed by the Security Domain}
\label{challenges}
While many of the usual obstacles of entity extraction, such as a lack of training data or semantic drift, are focal points of previous research, our initial efforts in the  cyber-security domain have illuminated challenges not often addressed in the literature.  

\begin{itemize}

\item {\bf Inaccuracies in source documents.} Often in online sources, entities are discussed with incorrect names.  
Particularly glaring examples are the erroneous, synonymous use of the terms ``malware'' and ``exploit'',  and the use of ``virus'' as  a blanket term for any malware.  
Consequently, any automatic labeling which derives the appropriate entity type from the contextual information will be confined to the accuracy of the text itself.  
Currently, we have focused on accurately labeling the documents, which may not necessarily contain correct information for these reasons.   

\item {\bf Alternate meaning of terms.}  Many of the names useful to cyber-security are comprised of common terms which take on a loaded meaning in our domain such as ``exploit'',  ``application'', or `` host''.  
As a primary mitigation for this, we've only used text sources that are on-topic and plan to automate this task via a decision classifier to ensure relevance when populating the corpus. More importantly, the PACE algorithm only learns patterns from known (seeded or learned) [entity, context] instances, allowing only those ``true'' occurrences of entity names to have input to the learning process.   

\item {\bf Complicated entity names.} Entities to be extracted in our setting are often nuanced and complicated in comparison to previous bootstrapping efforts.  
For example, in \cite{brin1999extracting, carlson2010coupled, carlson2010website} entity types are stereotypical proper nouns, such as Countries, Athletic Teams, Authors, or Book Titles, which are easier to identify than the more esoteric entities we desire, such as a Vulnerability Potential Effect, often because the latter entities occur as short phrases and are not as distinguishable by their syntactic features (e.g. always beginning with a capital letter).  
Our current effort has addressed this by chunking the text before pattern matching to ensure descriptive phrases are extracted.   
\end{itemize}

\section{Future Work \& Conclusion}
\label{future} 
While the theoretical foundation for PACE is set, tuning it to work well will require more work.  
Designing and testing a more robust scoring  for patterns, such as negatively scoring patterns which are observed with a variety of entity types,  should increase performance.  
While an advantage exhibited above  was the ability of PACE to operate on a small corpus, we expect PACE to excel in the presence of an enormous corpus. 
We aspire to test PACE against a traditional bootstrapping algorithm  with  common seeds and corpora to get true comparative results, in particular with respect to speed, and pattern accuracy.  In addition to increasing the corpus size, adding more seed examples should provide an immediate boost. 

Operationally, PACE will fit into a larger architecture  that  will collect data from the web and populate the corpus with documents that are first deemed relevant by a decision classifier.  We envision a document stream that allows documents to live in the corpus for a fixed time or number of cycles before being discarding; hence, the ability of PACE to work on various sized corpora is necessary for it to run continuously as the corpus is in flux.  
In addition, modifying PACE to extract relations is a next step in truly gaining knowledge from extracted concepts.  Development of an ontology to organize the entities into a graph is underway. Following the results of \cite{carlson2010coupled}, incorporating constraint conditions imposed by the ontology should increase accuracy as well.

Overall, we have provided the foundation of a semi-supervised tool, PACE, with makes novel contributions to the bootstrapping process for entity extraction.  Most notably, PACE strengthens pattern nomination and obviates a costly corpus search by storing contextual information along with seeded and promoted entities.  
Additionally, PACE has been shown to meaningfully extract desired entities in the presence of a relatively tiny corpus, which increases recall on any corpus document.  
Upon further development PACE should automatically extract relevant relations, and work in an architecture which allows fluid analysis of new documents.

\bibliographystyle{abbrv} % not sure what all the options are here
\bibliography{bootstrapping} % name of .bib file here
\end{document}